\documentclass[fleqn,usenatbib]{mnras}

\usepackage{newtxtext,newtxmath}

\usepackage[T1]{fontenc}
\usepackage{ae,aecompl}

\usepackage{graphicx}	
\usepackage{amsmath}	
\usepackage[dvipsnames]{xcolor}
\usepackage{multirow}
\usepackage{delarray}
\usepackage{color}
\usepackage{here}
\usepackage{float}
\usepackage{tensor}
\usepackage{xspace}
\usepackage{orcidlink}
\usepackage{nicefrac}
\usepackage[multiple]{footmisc}

\newcommand{\be}{\begin{equation}} \newcommand{\ee}{\end{equation}}

\newcommand{\solarmass}{M_{\rm \sun}}
\newcommand{\msun}{\solarmass}

\newcommand{\appropto}{\mathrel{\vcenter{
  \offinterlineskip\halign{\hfil$##$\cr
    \propto\cr\noalign{\kern2pt}\sim\cr\noalign{\kern-2pt}}}}}

\defcitealias{hopkins2015_gizmo}{H15}
\defcitealias{hopkins_gizmo_mhd}{HR16}
\defcitealias{Springel_2005_gadget}{S05}
\defcitealias{truelove_1997_dens_condition}{T97}
\defcitealias{guszejnov_isothermal_mhd}{Paper 0}
\defcitealias{grudic_starforge_methods}{Paper I}
\defcitealias{guszejnov_starforge_jets}{Paper II}

\usepackage[normalem]{ulem}

\title[Accelerating self-gravitating hydrodynamics]{Accelerating self-gravitating hydrodynamics simulations with adaptive force updates}

\author[M. Y. Grudi\'{c}]{
Michael Y. Grudi\'{c}\orcidlink{0000-0002-1655-5604}$^{1}$\thanks{mike.grudic@northwestern.edu}
\\
$^{1}${CIERA and Department of Physics and Astronomy, Northwestern University, 2145 Sheridan Road, Evanston, IL 60208, USA}\\
}

\date{\today \vspace{-0.6cm}}

\pubyear{2020}

\begin{document}
\label{firstpage}
\pagerange{\pageref{firstpage}--\pageref{lastpage}}
\maketitle 

\begin{abstract}
Many astrophysical hydrodynamics simulations must account for gravity, and evaluating the gravitational field at the positions of all resolution elements can incur significant cost. Typical algorithms update the gravitational field at the position of each resolution element every time the element is updated hydrodynamically, but the actual required update frequencies for hydrodynamics and gravity can be different in general. We show that the gravity calculation in hydrodynamics simulations can be optimised by only updating gravity on a timescale dictated by the already-determined maximum timestep for accurate gravity integration $\Delta t_{\rm grav}$, while staying well within the typical error budget of hydro schemes and gravity solvers. Our implementation in the {\small GIZMO} code uses the tidal timescale introduced in \citet{tidaltimestep} to determine $\Delta t_{\rm grav}$ and the force update frequency in turn, and uses the rate of change of acceleration evaluated by the gravity solver to construct a predictor of the acceleration for use between updates. We test the scheme on standard self-gravitating hydrodynamics test problems, finding solutions very close to the standard scheme while evaluating far fewer gravity forces, optimising the simulations. We also demonstrate a $\sim 70\%$ speedup in an example simulation of a giant molecular cloud. In general, this scheme introduces a new tunable parameter for obtaining an optimal compromise between accuracy and computational cost, in conjunction with e.g. time-step tolerance, numerical resolution, and gravity solver tolerance. 
\end{abstract}



\section{Introduction}
Many interesting astrophysical phenomena involve an interplay of hydrodynamics and gravity, and the vast majority of such problems resist analytic solutions. Hence numerical hydrodynamics simulations are an important tool for understanding self-gravitating astrophysical flows. Numerical hydrodynamics methods are necessarily approximate, only converging to the exact solution in the limit of infinite resolution (if even that), so whenever numerical hydrodynamics is adopted for a problem, one can only hope to obtain a simulation setup that achieves the desired solution accuracy for the least CPU time possible. The trade-off between cost and accuracy generally has multiple knobs one can turn, such as the numerical resolution and the accuracy of the gravity solver.

Evaluating the gravitational field $\mathbf{g}$ at the positions of all hydrodynamic elements can be a significant portion of the overall computational cost of a simulation. The na\"{i}ve algorithm scales as $N^2$, where $N$ is the number of resolution elements, which is impractical for all but the lowest-resolution simulations. Even more-efficient Fourier transform or multipole algorithms that scale $\propto N\log N$ \citep[e.g.][]{hockney_eastwood_pm, barneshut} or $\propto N$ \citep[e.g.][]{dehnen_2000_fmm, gadget4} can sometimes dominate over other workloads, such as the actual hydrodynamic evolution. This issue can be particularly pronounced in massively-parallel computations, in which the long-ranged nature of gravity makes it inherently communication-intensive compared to hydrodynamics (which only generally needs to handle local interactions with neighboring elements), incurring increasingly-large overheads as the size of the simulation is increased.

Most self-gravitating hydro codes update the gravitational field $\mathbf{g}$ with the gravity solver every time a resolution element is updated hydrodynamically, but the actual requirements for the update frequencies of hydrodynamics and gravity are not the same. The maximum time-step between hydro updates is typically constrained by the rate at which information propagates across the resolution elements, the \citet{cfl} (CFL) criterion:
\begin{equation}
    \Delta t_{\rm hydro} < C_{\rm CFL} \frac{\Delta x}{v_{\rm sig}},
\end{equation}
where $\Delta x$ is length of a hydrodynamic cell ($\sim \left(\Delta m/\rho\right)^{1/3}$ in the case of Lagrangian discretizations with a certain mass resolution $\Delta m$, at a given density $\rho$), $v_{\rm sig}$ is an estimate of the maximum velocity at which information propagates across the cell, and $C_{\rm CFL}$ is a constant chosen to ensure stability for a given solver method. 

In contrast, the required frequency of gravity updates is generally set by the requirement of accurate orbital integration \citep[e.g.][]{dehnen_nbody_review}:
\begin{equation}
    \Delta t_{\rm grav} < \sqrt{\eta} t_{\rm dyn},
\end{equation}
where $\eta$ is a tolerance parameter set to give the desired orbital integration accuracy and $t_{\rm dyn}$ is some estimate of the local gravitational dynamical timescale, nominally of order $\sqrt{\frac{R^3}{G M}}$ where $R$ and $M$ are the characteristic size and mass of the system. This does not necessarily have anything to do with the simulation resolution, and need not approach 0 in the limit of infinite resolution, if a certain error in orbital integration accuracy is to be tolerated. And in many practical cases, an error in the orbital integration exists {\it regardless} of the time-step chosen, because an approximate method must be used to compute the gravitational acceleration $\mathbf{g}$ for the calculation to be tractable. Note that in the case of hydrostatic equlibrium, the global sound crossing time and $t_{\rm dyn}$ are of comparable magnitude

Hence, the required update frequencies for stable integration of gravitational motion and stable hydrodynamic evolution can be quite different in practice. It follows that there exist regimes in which the time-step demanded by hydrodynamics (or other physics, such as MHD, chemistry, or radiation) is more stringent than that required for gravity. In such instances, it may not be necessary to update the gravitational field every single hydro time-step, because there is a separation of timescales between the local hydrodynamic evolution and orbital motion. Because the variation in  $\mathbf{g}$ is ``smooth" on the hydrodynamic timescale, it is possible to accurately predict the evolution of $\mathbf{g}$ between hydrodynamic time-steps without having to directly evaluate it from the current mass distribution in the simulation each time. If evaluating $\mathbf{g}$ is a significant computational cost (often the case in large simulations), such a scheme could reduce the cost of the simulation overall, or allow more-accurate simulations at fixed cost.

In this paper we describe such a scheme for adaptively updating the gravitational field in hydrodynamics simulations, which uses the gravitational timestep criterion $\Delta t_{\rm grav}$ to determine how often a new gravity calculation is actually needed for a given resolution element, and uses a predictor to update $\mathbf{g}$ between hydrodynamic time-steps, potentially achieving comparable accuracy at greatly-reduced cost. In \S\ref{sec:method} we describe the scheme for adaptively updating $\mathbf{g}$ and predicting it between updates. In \S\ref{sec:tests} we describe an implementation of the scheme in the {\small GIZMO} code and test it on a variety of gravitational collapse problems where accurate coupling of hydrodynamics and gravity are important, demonstrating that it can achieve comparable accuracy at reduced cost compared to the na\"{i}ve algorithm. In \S\ref{sec:conclusion} we summarize our findings and discuss various caveats and possible extensions to our approach.

\section{Algorithm}
\label{sec:method}
Consider any simulation setup that solves at least the equations of self-gravitating hydrodynamics (with possible additional physics as well) by updating hydrodynamic quantities in discrete timesteps. Let $\Delta t_i$ be the time-step that hydro resolution element $i$ is assumed to take, generally set by the combined requirements for stably integrating all physics included in the simulation -- at least gravity and hydrodynamics, but also potentially magnetic fields, diffusion, radiation, etc. As mentioned in the introduction, $\Delta t_i$ will generally be chosen to ensure some baseline integration accuracy for gravitational motion, but can often be overruled by more-stringent requirements from other physics, i.e. $\Delta t_i = \min \left(\Delta t_{\mathrm{grav},i}, \Delta t_{\mathrm{hydro},i} \right)$\footnote{Other interpolations between different timestep criteria can be used, e.g. the harmonic mean instead of the minimum, but in general timestepping schemes must  adopt some function that limits the timestep to the most stringent of all timestep requirements.}, where $\Delta t_{\mathrm{grav},i}$ and $\Delta t_{\mathrm{hydro},i}$ are the respective timesteps required to ensure stablility/accuracy for gravity and all other physics (but at least hydrodynamics) respectively. In standard schemes, the gravitational field $\mathbf{g}_i$ is updated every $\Delta t_i$, according to the chosen gravity solver. The technique described here is not specific to any particular gravity solver, but obviously the potential speed-up depends upon the cost of running the solver compared to the rest of the simulation. 


\subsection{Adaptive force updates}
Whenever $\mathbf{g}_i$ is updated by the gravity solver, initialize a counter $T_{\mathrm{f},i}$ to zero, specific to the hydro resolution element $i$. This counter will track the time elapsed since the last force update, i.e. at the beginning of every subsequent timestep we update it as
\begin{equation}
    T_{\mathrm{f},i} \mapsto T_{\mathrm{f},i}  + \Delta t_i.
\end{equation}
Then, our adaptive force updating strategy is to only update $\mathbf{g}_i$ once at least a certain amount of time $q_\mathrm{f} \Delta t_{\mathrm{grav},i}$ has passed, i.e.
we wait to re-evaluate $\mathbf{g}_i$ with a new force computation until the condition
\begin{equation}
    T_{\mathrm{f},i} \geq q_\mathrm{f} \Delta t_{\mathrm{grav}, i},
    \label{eq:updatecondition}
\end{equation}
is satisfied, where $q_\mathrm{f}$ is a chosen numerical tolerance parameter. Once this is done, we reset $T_{\mathrm{f},i}$ to 0 and start the cycle anew. Hence we only update gravity with a frequency related by a factor $1/q_\mathrm{f}$ to the time-stepping frequency required for integrating gravitational motion. $q_\mathrm{f}$ can conceivably take any value, with larger values requiring less-frequent updates and smaller values ensuring more-frequent updates, with corresponding trade-offs in accuracy. It is easily seen that for $q_\mathrm{f}\leq 1$ this method reverts exactly to the normal behaviour of the code in the regime where $\Delta t_i=\Delta t_{\mathrm{grav},i}$ -- specifically it would only change the code behaviour when $\Delta t_i = \Delta t_{\mathrm{hydro},i} < q_\mathrm{f} \Delta t_{\mathrm{grav},i}$.


\subsection{Adding a predictor}
We could stop here and simply accept the error term arising from less-frequent updates with no other modifications, but the estimate of $\mathbf{g}_i$ we would use has a leading-order error term due to the systematic time lag with respect to the true $\mathbf{g}_i$, which may be more likely to compound over time than e.g. the more randomly-distributed errors often introduced by approximate gravity solvers. This term can be eliminated fairly easily with a {\it predictor} for updating $\mathbf{g}_i$ between force evaluations. One possibility is a simple polynomial predictor, obtained by logging some number of past $\mathbf{g}_i$ values and using an interpolating function to predict future $\mathbf{g}_i$ values between gravity solves, with a first-order (linear) polynomial being sufficient to eliminate the leading-order time lag error. Alternatively, for some gravity solvers it may be convenient to directly evaluate the gravitational jerk $\mathbf{j}_i$:

\begin{equation}
    \mathbf{j}_i\coloneq \frac{\rm d}{\rm d t} \mathbf{g}_i = G \int \mathrm{d}^3 \mathbf{x}\, \rho(\mathbf{x}) \left(\frac{\mathbf{v}(\mathbf{x})-\mathbf{v}_i}{r^3}- 3 \frac{\left[\left(\mathbf{v}(\mathbf{x})-\mathbf{v}_i\right)\cdot \mathbf{r}\right]\mathbf{r}}{r^5}\right),
\end{equation} 
where $\rho(\mathbf{x})$ is the mass density, $\mathbf{r}=\mathbf{x}-\mathbf{x}_i$ and $\mathbf{v}\left(\mathbf{x}\right)$ is the average velocity integrated over the phase-space distribution at point $\mathbf{x}$, and naturally the integral should be discretized according to a quadrature rule consistent with the discretization of the hydro method. If $\mathbf{j}_i$ is available, we can update $\mathbf{g}_i$ according to a first-order predictor
\begin{equation}
    \mathbf{g}_i \mapsto \mathbf{g}_i + \mathbf{j}_i \Delta t_i
    \label{eq:jerkupdate}
\end{equation}
between force updates from the gravity solver.

\section{{\small GIZMO} implementation and tests}
\label{sec:tests}
We have implemented a version of the scheme described in \S\ref{sec:method} in the {\small GIZMO} code \citep{hopkins2015_gizmo}. {\small GIZMO} solves the equations of self-gravitating hydrodynamics (among many other optional physics) with a variety of available methods, including the mesh-free finite-volume Meshless Finite-Volume (MFV) and Meshless Finite-Mass (MFM) methods \citep{gaburov_nitadori_mfv, hopkins2015_gizmo}, various flavours of smoothed particle hydrodynamics (SPH, \citealt{Hopkins_SPH_2013}), and finite-volume Eulerian or moving-mesh methods in Cartesian, spherical, and cylindrical coordinates. All tests presented here use the MFM solver, but our implementation of the scheme presented here is not specific to any one solver.

{\small GIZMO} solves gravity with an approximate oct-tree-based tree-code solver \citep{barneshut,Springel_2005_gadget}, optionally in combination with an approximate particle-mesh FFT solver to handle long-ranged forces. Details of the massively-parallel algorithms for each solver, and the specifics of the long/short-range force decomposition, are given in \citet{Springel_2005_gadget}. All tests considered here use only the tree-code solver, including additional \citet{ewald} summation terms in problems with periodic boundary conditions. Following \citet{grudic_2020_cluster_formation}, we supplement the tree-node opening criteria given in \citet{Springel_2005_gadget} with the original \citet{barneshut} geometric criterion, such that a node of side-length $L$ is always split if it subtends an angle $L/r > \Theta$, where $\Theta=0.5$ is the maximum opening angle.  We adopt fully-adaptive gravitational softening in all problems (i.e. no minimum or maximum softening length), maintaining consistency between the gravitational and hydrodynamic resolution everywhere in the simulation. We include the additional force terms required for energy conservation in conjunction with adaptive softening \citep{price_2007_adaptive_softening}.

{\small GIZMO} uses an adaptive, hierarchical powers-of-two individual block timestepping scheme \citep{Springel_2005_gadget}. Our implementation of the adaptive force updating scheme in \S\ref{sec:method} is conceivably compatible with any estimator of $\Delta t_{\rm grav}$ (e.g. the standard acceleration-based criterion, \citealt{power_2003_timestep}), but here we use the tidal timestep criterion introduced in \citet{tidaltimestep}:
\begin{equation}
    \Delta t_{\mathrm{grav},i} = \Delta t_{\mathrm{tidal},i} = \sqrt{\eta}\left(\frac{\|\mathbf{T}_i\|^2}{6}\right)^{-1/4},
    \label{eq:tidaltimestep}
\end{equation}
where $\mathbf{T}_i$ is the tidal tensor evaluated at the position of element $i$, $\|\cdot\|$ is the Frobenius norm, and $\eta$ is a tolerance parameter setting the time integration accuracy (all tests here set $\eta=0.01$). In \citet{tidaltimestep} we found this to have good performance as a general-purpose timestep criterion for multi-physics simulations with gravity, because the tidal timescale $t_{\rm tidal} \equiv \left(\|\mathbf{T}\|^2/6\right)^{-1/4}$ is a good proxy for the gravitational dynamical time of the system $t_{\rm dyn} \sim \Omega^{-1} \sim \sqrt{r^3 / G M(<r)}$. Hence it is also useful as an estimate of how often a new force evaluation is needed.

We emphasize that adaptive force updating does {\it not} generally require any particular choice of $\Delta t_\mathrm{grav}$ or predictor scheme, and in general could be implemented without the additional overhead of evaluating the tidal tensor or the jerk. Our approach is mainly motivated by use cases that are computing these quantities anyway, e.g. multi-physics star formation simulations \citep{starforge_methods, guszejnov_starforge_jets}. However the relation between $q_\mathrm{f}$, the error, and the computational speedup will be specific to a particular implementation: other approaches may need to re-tune $q_\mathrm{f}$ to achieve errors/speedups similar to what we present here, and may achieve worse or better speedups at fixed error.

To validate the method and build some intuition for the trade-off between accuracy and efficiency parametrized by $q_\mathrm{f}$, we consider the performance of this setup on a variety of test problems.

\subsection{Adiabatic collapse}
\begin{figure}
    \centering
    \includegraphics[width=\columnwidth]{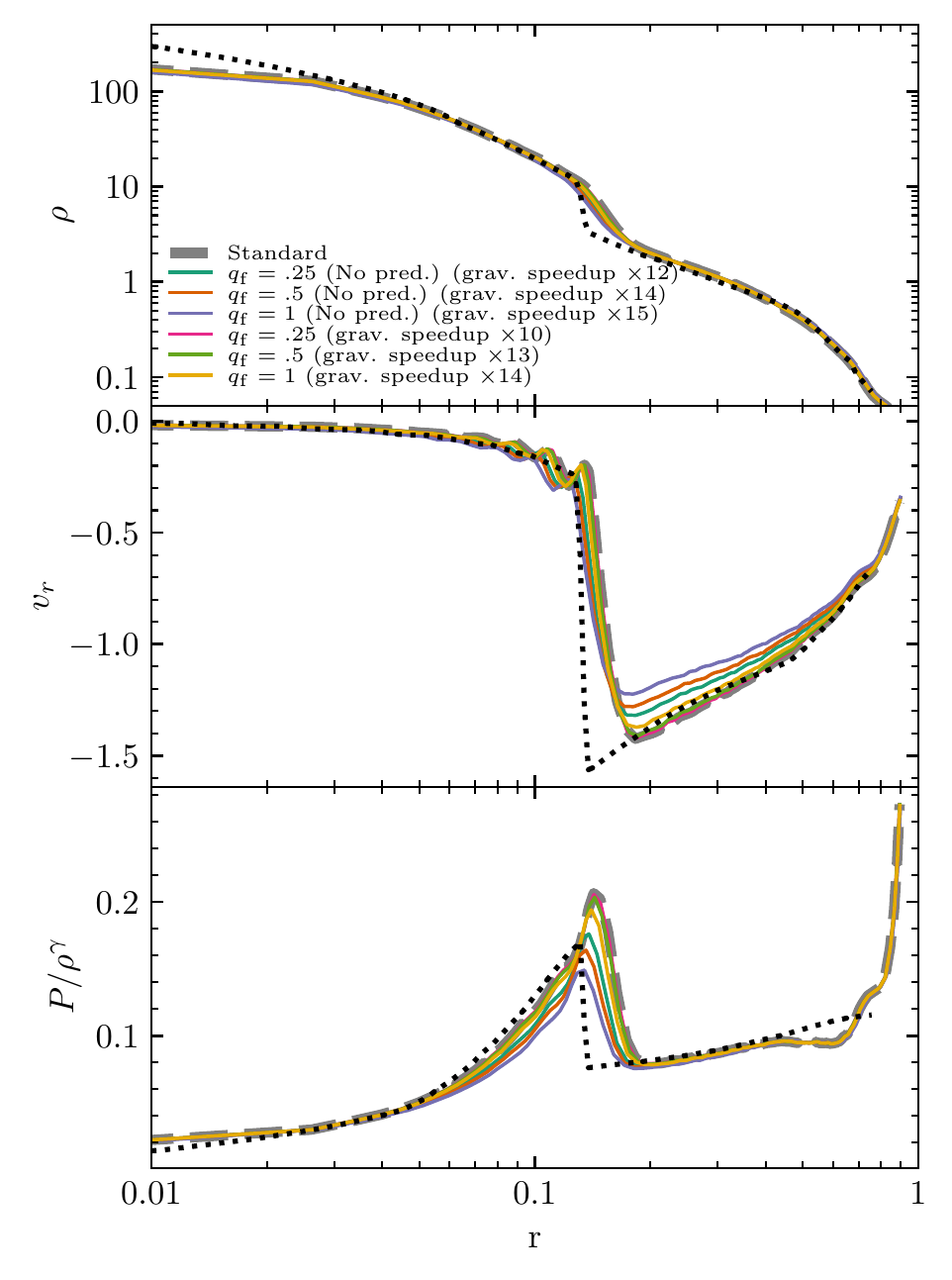}\vspace{-6mm}
    \caption{Radial profiles of density $\rho$, radial velocity $v_r$, and the entropic function $P/\rho^\gamma$ in the \citet{evrard_test} spherical adiabtic collapse test run in 3D with $30^3$ gas cells with the {\small GIZMO} MFM method. We compare results at $t=0.9$ for a range of $q_\mathrm{f}$ settings to the standard test with no adpative force updates, as well as a more-accurate solution from a much higher-resolution 1D PPM calculation (dotted, \citealt{steinmetz_evrard_ppm}). ``No pred." runs neglect the first-order predictor otherwise used between force updates (Eq. \ref{eq:jerkupdate}). Note that the {\it overall} speedup is small here because gravity is never the dominant cost of this low-resolution simulation.}
    \label{fig:evrard}
\end{figure}

We start with the \citet{evrard_test} spherical adiabatic collapse problem, a popular test problem with a well-accepted solution that has been run with many codes \citep{hernquist_katz_sph,steinmetz_evrard_ppm, dave_1997_treesph,gasoline,Springel_2005_gadget, arepo, hopkins2015_gizmo}. The initial condition is an isolated gas sphere of unit mass and radius with density profile $\rho \propto r^{-1}$ truncated at $R=1$, initially at rest with an initial specific internal energy of $0.05$ and an adiabatic index $\gamma=5/3$. Following \citet{hopkins2015_gizmo} we realize this initial condition in the MFM method in $30^3$ fixed-mass Lagrangian gas cells. The solution consists of an initial gravitational collapse that proceeds until a central ``bounce" occurs and a shock plows through subsequent infalling material. We run a standard benchmark test without adaptive force updating, and survey a range of $q_\mathrm{f}$ between $1/4$ and $1$ with the scheme. We also run respective versions without the jerk predictor (Eq. \ref{eq:jerkupdate}) to verify that it is eliminating an error term, for a total of one benchmark test, three tests with the full scheme, and three tests without the predictor.

In Figure \ref{fig:evrard} we plot the radial profiles of density $\rho$, radial velocity $v_r$, and the entropic function $P/\rho^\gamma$ at $t=0.8$ for the different runs, compared to the result of a much higher-resolution 1D simulation run with the piecewise-parabolic method (PPM, \citealt{steinmetz_evrard_ppm}), which can be treated as the ``exact" solution for the present purposes. In Fig. \ref{fig:evrard} we also quote the respective speedups for gravity operations. We first note that the predictor {\it can} reduce the error considerably: the result of the run without a predictor with $q_\mathrm{f}=1/4$ (more-frequent updates) had greater errors with respect to the standard run than the run with the predictor with $q_\mathrm{f}=1$ (less-frequent updates). We use the predictor in all subsequent tests. The predictor runs with smaller $q_\mathrm{f}$ generally had errors $\lesssim 1\%$ with respect to the standard run, for all quantities, despite the fact that {\it all} adaptive runs evaluated at least ten times fewer gravity forces! The {\it overall} speedup here is negligible because this low-resolution run is bottlenecked by the hydrodynamics rather than gravity, but other simulations can be more gravity-bound, as in our subsequent tests.

It should also be noted that {\it all} of the adaptive solutions are much more similar to the standard run than the standard run is to the exact solution, hence in these instances the gravity error introduced by adaptive updates is not the dominant error. Rather, the errors lie mainly in the hydrodynamics, and formally converge away with sufficient hydrodynamic resolution in this problem \citep{hopkins2015_gizmo}. As long as hydro errors do dominate, it is more optimal to invest compute time running simulations at higher resolution than it is to ensure highly-accurate gravity at fixed resolution.

\subsection{Interacting polytropes}
\label{sec:polytrope}
\begin{figure}
  \includegraphics[width=\columnwidth]{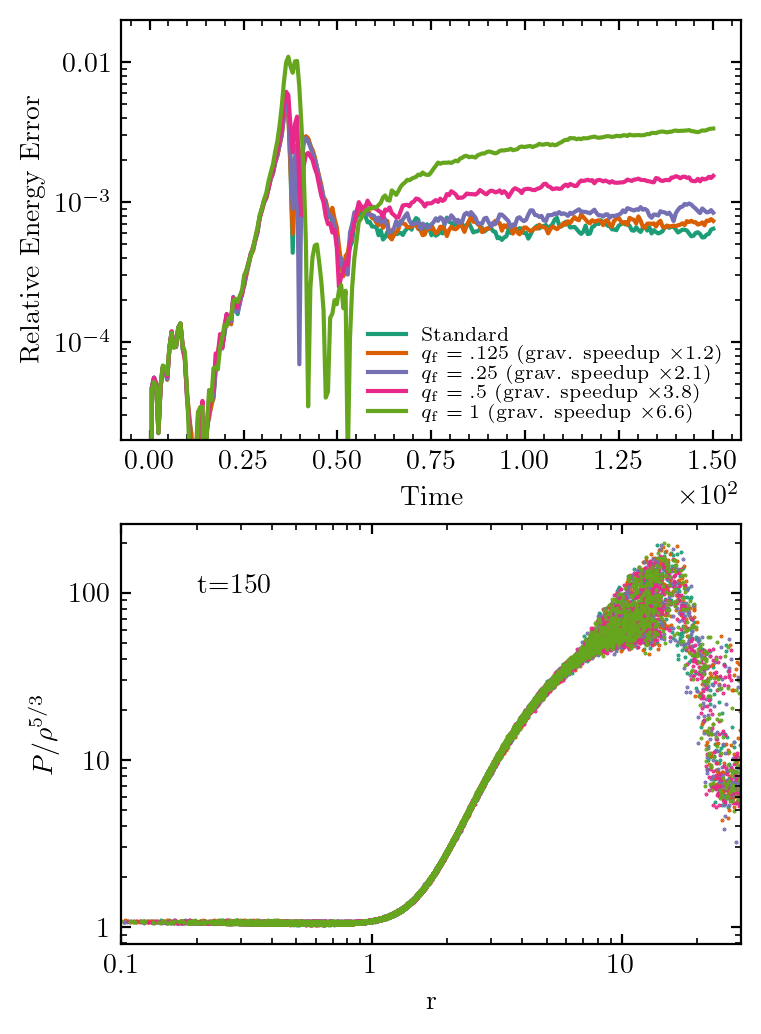}\vspace{-6mm}
  \caption{Results of the \citet{nelson1994} merging $n=3/2$ polytropes test. {\it Top}: Evolution of the relative energy error for the standard run and different adaptive force update tolerances $q_{\rm f}$. {\it Bottom}: Radial entropy profile of the final merger product.}
  \label{fig:polytrope}
\end{figure}
As a second test of self-gravitating adiabatic hydrodynamics, we simulate the dynamics of two identical $n=3/2$ polytropes evolved with a $\gamma=5/3$ equation of state in a head-on merger from rest at an initial separation of $6\times$ their radius, following \citet{nelson1994}. The two polytropes (with $10^4$ gas cells each) fall into each other and collide, forming a central shock, and eventually the cores merge and a new ``star'' is formed (with some residual expelled material). This tests the performance of the updating scheme both in the highly-dynamical regime of the approach and merger, and during the near-hydrostatic ringdown of the merger product.

Physically, energy is conserved in this problem, so we examine the evolution of the total (kinetic, thermal, and gravitational) energy $E_{\rm tot}$ to quantify the simulation error as a function of $q_{\rm f}$, with $q_{\rm f}$ ranging from $0.125-1$. In the top panel of Figure \ref{fig:polytrope} we plot the evolution of the total energy error through the merger and ringdown, and find that $q_{\rm f}$ does influence the error: greater values tend to produce greater errors. However, the incremental error is very small for $q_{\rm f} \lesssim 0.25$, and the {\it maximum} error is only notably greater for $q_{\rm f}=1$. In the bottom panel we also plot the final entropy profile of the merger product, and find no clear systematic difference in this result when varying $q_{\rm f}$.

Momentum should also be conserved in this problem, and one might also worry that by introducing errors in $\mathbf{g}$ with the force updating scheme one might introduce global momentum conservation errors. However, we did not find a systematic trend in the net momentum error with $q_{\rm f}$ in these simulations, so it appears that the momentum error introduced by the force updating scheme is subdominant to that introduced by the approximate Barnes-Hut gravity solver. We have also verified that $q_{\rm f}$ does not systematically increase momentum errors when the polytropes are instead evolved on a circular binary orbit.

\subsection{Isothermal collapse}
We now consider two variants of the \citet{boss_bodenheimer_1979} ``standard isothermal test case", following the collapse of a rotating, isothermal cloud with an initial density perturbation, which are useful for testing the ability of the scheme to follow the runaway collapse of gas over many orders of magnitude in density.

\subsubsection{\citet{burkert_bodenheimer_1993} version}
\begin{figure}
    \centering
    \includegraphics[width=\columnwidth]{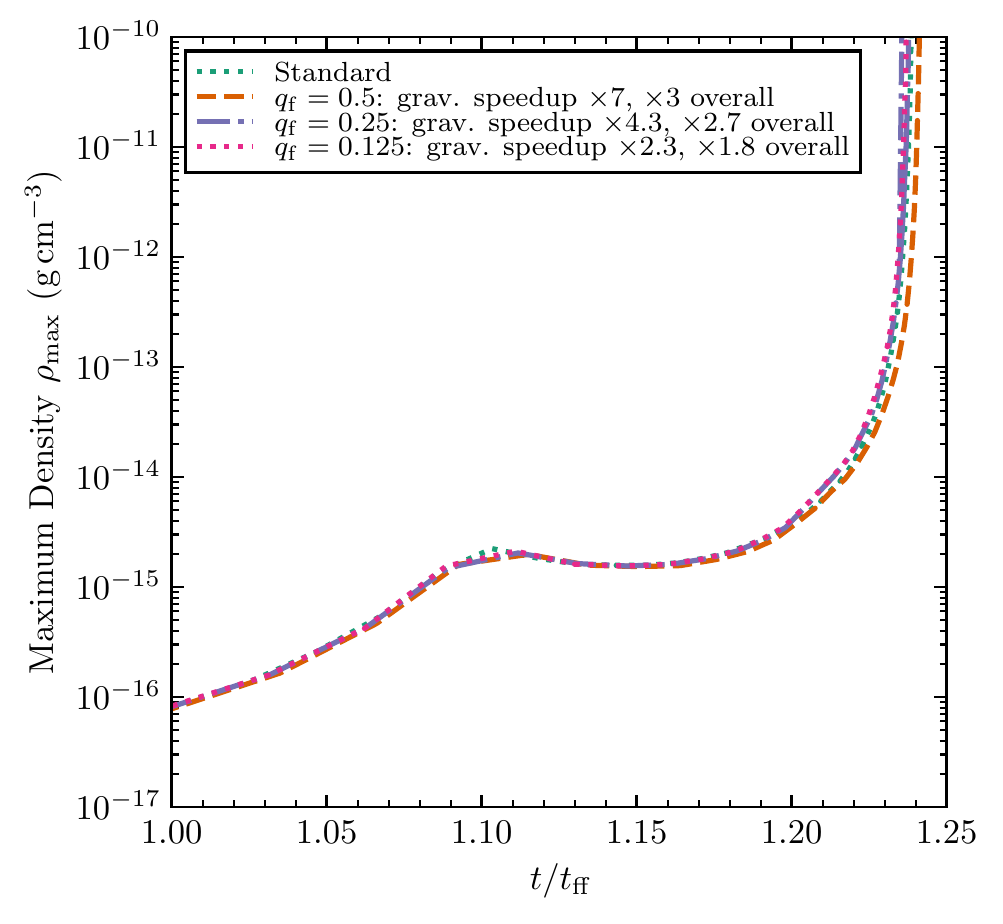}\vspace{-6mm}
    \caption{Maximum density as a function of time in the \citet{burkert_bodenheimer_1993} isothermal collapse problem, with time in units of the initial cloud freefall time, comparing a standard run with adaptive force updating runs with $q_\mathrm{f}=1/8-1/2$.}
    \label{fig:bossbodenheimer}
\end{figure}
This version consists of a uniform-density, un-magnetized, spherical solar-mass gas core with initial radius $5\times 10^{16}\rm cm$, in uniform rotation with $\Omega = 7.2 \times 10^{-13} \rm rad \,s^{-1}$ and a $10\%$ $m=2$ azimuthal density perturbation \citep{burkert_bodenheimer_1993, bate_1997_resolution, Springel_2005_gadget}. We employ an isothermal equation of state $P=c_{\rm s}^2 \rho$, $c_{\rm s}=0.166\rm km\,s^{-1}$, and initialize the cells in a uniform-density glass configuration with the density perturbation imposed by rescaling the cell masses slightly. All tests resolve the cloud in $10^7$ gas cells, for an average mass resolution of $10^{-7} \msun$.

In Figure \ref{fig:bossbodenheimer} we plot the evolution of the maximum density in the simulation as a function of time, in units of the initial cloud freefall time $t_\mathrm{ff} = \sqrt{3\uppi/\left(32 G \rho\right)}$. We find that the results are difficult to distinguish from the standard solution for the entire range of $q_\mathrm{f}$ values we have considered, $1/8-1/2$, with the exception of the late-time behaviour of the $q_\mathrm{f}=1/2$, which experiences just a slight delay in collapse. This cost of this simulation is normally dominated by gravity, so all adaptive runs gave a significant overall speedup, ranging from $1.8-3$.

\subsubsection{\citet{truelove_1997_dens_condition} version}
\begin{figure*}
\includegraphics[width=\textwidth]{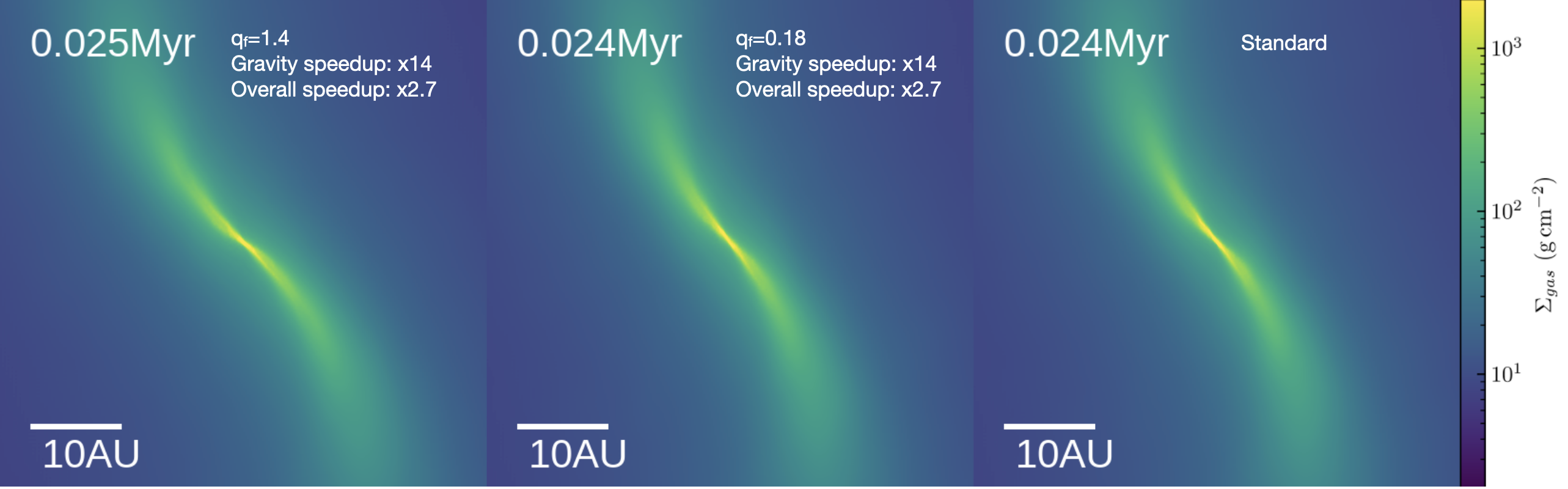}\vspace{-6mm}
\caption{Gas surface density maps in the \citet{truelove_1997_dens_condition} isothermal collapse problem when the maximum density reaches $10^{-9.5}\rm g \, cm^{-3}$, for varying force accuracy parameters $q_\mathrm{f}=1.4$ ({\it left}), $q_\mathrm{f}=0.18$ ({\it centre}) and the reference run without adaptive force updates ({\it right}). We quote the respective speedups of gravity and the overall simulation compared to the standard run.} 
\label{fig:truelove}
\end{figure*}
We also consider the version of this problem with an initial Gaussian density profile, also initially in solid-body rotation with a $10\%$ $m=2$ initial density perturbation, as originally studied by \citet{truelove_1997_dens_condition}. This is also a useful test because the {\it qualitative} nature of the solution has been confirmed by multiple codes with different methods \citep[e.g. see also][]{kitsionas_whitworth_sph_splitting, starforge_methods}: the converged result is a single filament that collapses toward its axis in finite time \citep{Inutsuka_Miyama_1992}, and this result can be sensitive to numerical resolution and errors in the the coupling of hydrodynamics and gravity (e.g. exposing artificial fragmentation if a certain method subject to it). We simulate this problem with $6.4\times 10^6$ gas cells, for a mean mass resolution of $1.57\times 10^{-7} \msun$.

In Figure \ref{fig:truelove} we plot the structure of the filament at the time at which the maximum density is  $10^{-9.5}\rm g \, cm^{-3}$, comparing the results of runs with $q_\mathrm{f}=1.4$ and $q_\mathrm{f}=0.18$ with the standard run, and quote respective speedups for gravity and overall. The $q_\mathrm{f}=1.4$ run collapsed slightly ($\sim 1 \rm kyr$) later, and picked up a slight phase error in the orientation of the filament, but is otherwise qualitatively similar to the standard solution. The $q_\mathrm{f}=0.18$ run is difficult to distinguish from the standard run, both morphologically in Fig. \ref{fig:truelove} and in every other metric that we have checked (time evolution of density statistics, gravitational energy, and kinetic energy). Both adaptive runs had significant overall speedups, running more than a factor of 2 faster due to the greatly-reduced gravity load. Here the $q_\mathrm{f}=0.18$ run is likely near the ``sweet spot" for the compromise between accuracy and cost, because it enjoyed most of the overall speedup obtainable by going to larger $q_\mathrm{f}$, but was also highly accurate with respect to the reference solution.

\subsection{Cosmological collapse: the Zeldovich Pancake}
\begin{figure}
    \centering
    \includegraphics[width=\columnwidth]{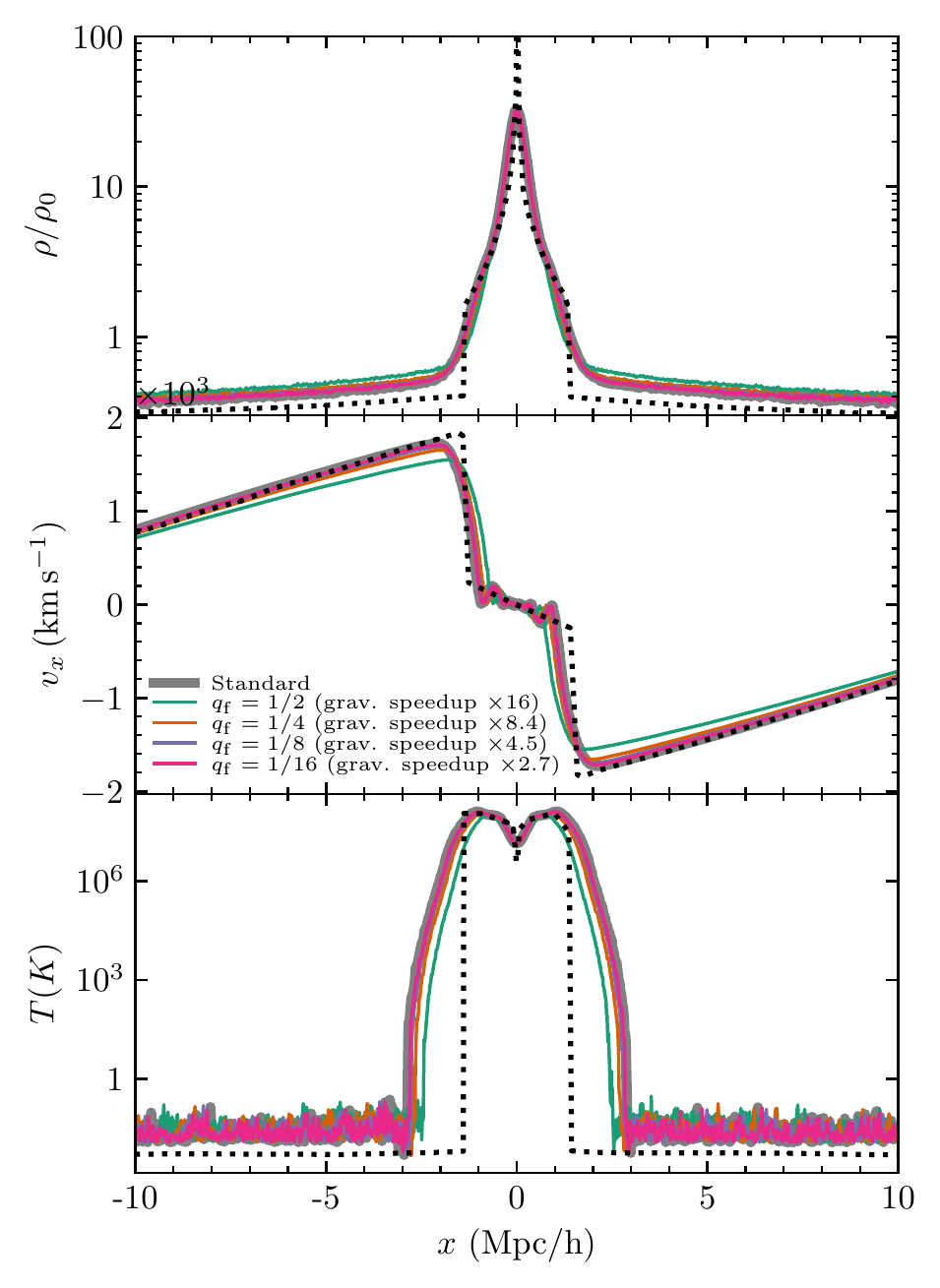}\vspace{-6mm}
    \caption{Results of the ``Zeldovich Pancake" test problem at $z=0$, plotting the gas density ({\it top}), velocity ({\it middle}), and temperature ({\it bottom}) as a function of position along the $x$ axis, for a variety of settings of gravity update frequencies $q_\mathrm{f}$. We compare with a solution from a 1D PPM code with much higher resolution, which should approximate the exact solution (dotted).}
    \label{fig:zeldovich}
\end{figure}

To test the the method in a cosmological scenario (e.g. for cosmological galaxy formation simulations), we simulate the ``Zeldovich Pancake" \citep{zeldovich1970}: the collapse of a single sinusoidal Fourier density mode in a periodic box in an Einstein-de Sitter universe. From the initial perturbation, the mode collapses to an eventual caustic at $z=1$.  We simulate this problem from $z=100$ to $z=0$ in a box of comoving length $64 \rm Mpc/h$ with $64^3$ gas cells (see \citet{hopkins2015_gizmo} for full details of the initial setup). We note that neither of {\small GIZMO}'s techniques for solving gravity with periodic boundary conditions (Ewald summation and particle-mesh) currently supports the jerk calculation, so only the contribution from the nearest periodic image of a mass element is accounted for, possibly reducing accuracy compared to non-periodic simulations.

In Figure \ref{fig:zeldovich} we plot profiles of overdensity, velocity, and temperature, comparing again with the result of a much higher-resolution (4096 cells across the box) PPM calculation. We survey $q_\mathrm{f}$ values between $1/16$ and $1/2$, only finding significant deviations in any of the plotted quantities for the $q_\mathrm{f}=1/2$ run. And even then, we again note that all of our runs are much more similar to the standard run than the standard run is to the ``exact" solution, indicating again the adaptive scheme is not the main contributor to the error. Overall speedups ranged from $1.4-6$: this problem is totally dominated by gravity costs because of the additional expense of Ewald summation and the high required force accuracy tolerances at high $z$, when the density field is nearly homogeneous \citep[e.g.][]{Springel_2005_gadget}.
\begin{figure}
    \centering
    \includegraphics[width=\columnwidth]{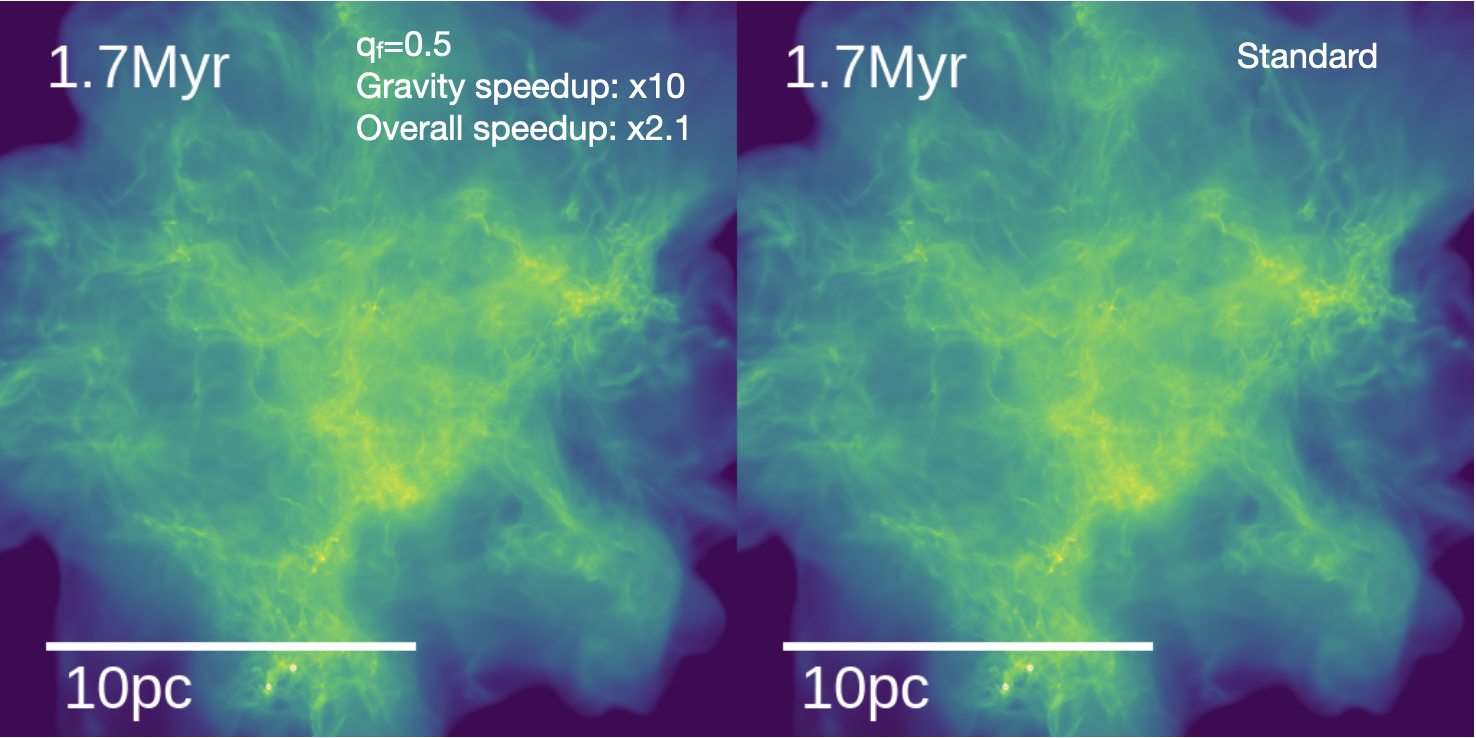}\vspace{-6mm}
    \caption{Comparison of the surface density morphology of a STARFORGE MHD GMC simulation run with and without adaptive force updates (with $q_\mathrm{f}=1/2$ in the adaptive run), shortly after the first stars have started forming. The morphologies are nearly identical, but the adaptive run had to spend only 1/10 as much CPU time in the gravity solver (and 1/2 as much CPU time overall) to reach this point.}
    \label{fig:starforge_morphology}
\end{figure}

\begin{figure*}
    \centering
    \includegraphics[width=\textwidth]{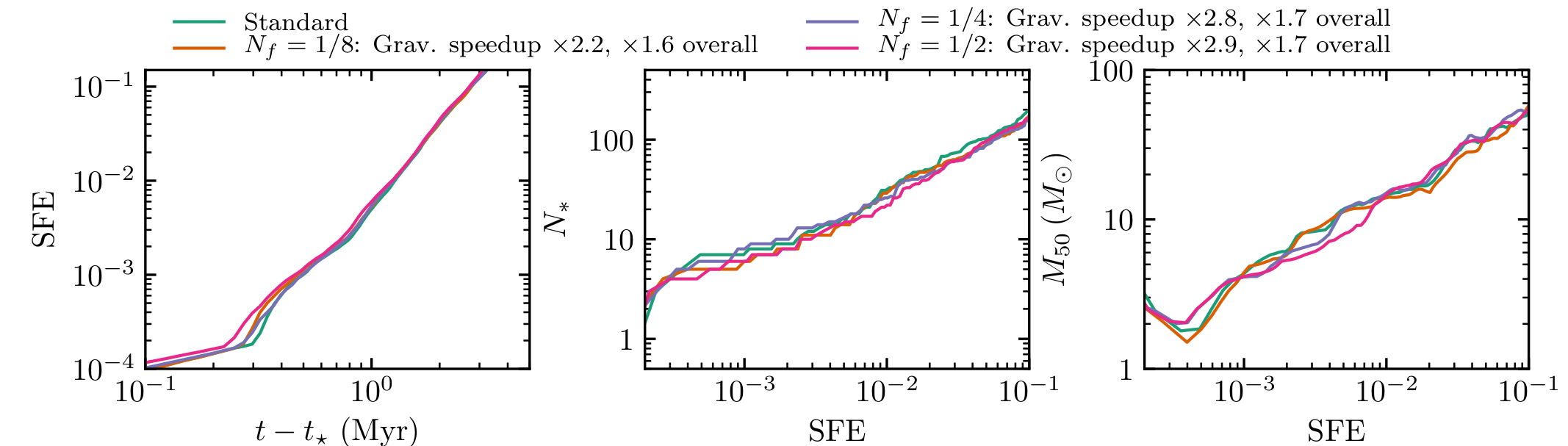}\vspace{-6mm}
    \caption{Time evolution of the star formation efficiency (SFE), and stellar mass function statistics measured at fixed SFE in an isothermal MHD simulation of a $2\times 10^4\msun$ GMC (\ref{sec:starforge}), comparing a standard, non-adaptive run with different adaptive force update intervals $q_\mathrm{f}$. The SFE is plotted as a function of time after the initial star formation. Stellar mass function statistics include the number of stars $N_\star$ and the mass-weighted median stellar mass $M_\mathrm{50}$.}
    \label{fig:starforge}
\end{figure*}

\subsection{Collapse of a turbulent magnetized GMC}
\label{sec:starforge}

Finally, we simulate a highly non-linear, chaotic, inherently three-dimensional self-gravitating MHD problem: the collapse and fragmentation of an isothermal, magnetized giant molecular cloud (GMC) \citep{price_bate_2007_mhd_sf,Haugbolle_Padoan_isot_IMF}, using the STARFORGE numerical framework \citep{starforge_methods} to follow the collapse, accretion, and dynamics of individual protostars represented by sink particles. Unlike previous problems in this section, this problem does not have an unambiguous or universally agreed-upon answer, but it is still useful to investigate the effect of $q_\mathrm{f}$ upon the main quantities of interest in star formation simulations: the star formation history, as expressed e.g. by the variation in the star formation efficiency (SFE) over time, and the statistics of the stellar mass function. 

We re-run the {\bf M2e4} GMC model introduced in \citet{guszejnov_isothermal_mhd}: an initially spherical GMC of uniform density with mass $2\times 10^4\msun$ and radius $10 \rm pc$, surrounded by a diffuse medium with $1/1000$ the density, threaded by a uniform initial magnetic field normalized so that the initial magnetic energy within the cloud itself is $1\%$ of the cloud's gravitational binding energy. The initial velocity field in the cloud consists of a Gaussian random field with power spectrum $P\left(k\right) \propto k^{-2}$, with a natural mixture of solenoidal and compressive modes, normalized so that the initial kinetic energy of the cloud equals the binding energy (virial parameter $\alpha_\mathrm{turb}=2$). We adopt a mass resolution of $10^{-3} \msun$, for a total of $2\times 10^7$ gas cells in the cloud. We enforce an isothermal equation of state with $c_\mathrm{s}=0.2\rm km\,s^{-1}$, and neglect all feedback mechanisms: although feedback can be important for achieving a realistic IMF \citep{Cunningham_2018_feedback,guszejnov_starforge_jets}, by allowing gravity to go unopposed we seek to intentionally exaggerate the effects of the gravity errors that we might introduce with adaptive force updates. We survey $q_\mathrm{f}$ values ranging from $1/8$-$1/2$, and also run a reference run with exactly the same code version. We refer the reader to \citet{starforge_methods} for all additional numerical details.

The GMC forms multiple self-gravitating cores through turbulent fragmentation \citep[e.g.][]{padoan_nordlund_1997_imf, padoan_nordlund_2011_imf,excursion_set_ism}, which undergo runaway collapse and are replaced by accreting sink particles (representing individual stars), which eventually accrete $10 \%$ of the total cloud mass by the end of the simulation. In Figure \ref{fig:starforge_morphology} we plot surface density maps of the $q_\mathrm{f}=1/2$ run and the standard run after the first two stars have formed after $1.7 \rm Myr$; their morphologies appear indistinguishable at this time, but the adaptive run has evaluated roughly 10 times fewer gravity forces. 

In Figure \ref{fig:starforge} we plot the evolution of the SFE as a function of time elapsed from the initial star formation, and find that the different runs are nearly indistinguishable from the reference run. We also compare various statistics of the stellar mass distribution at fixed SFE (see Fig \ref{fig:starforge} caption for full explanation), and although there exist visible fluctuations between the different runs (to be expected in such a non-linear problem subject to e.g. N-body chaos), the chosen value of $q_\mathrm{f}$ does not appear to influence the predicted IMF statistics in any clear, systematic way. Hence adaptive force updating is unlikely to influence any conclusions regarding the star formation history of the GMC or the roles of different physics in setting the IMF, the main science questions of this type of simulation.

Gravity speedups in the adaptive runs ranged from 2.2 to 2.9, and the overall speedups from 1.6 to 1.7. The savings on the cost of evolving {\it gas} is more significant, but our scheme does not eliminate the need to frequently update sink particles on short timesteps in hard binaries, which can often be the bottleneck for these simulations. Greater speedups may yet be possible in larger runs: the version of this run with $10\times$ higher mass resolution ran roughly $3\times$ faster overall with $q_{\rm f}=1/8$ than the version in \citet{guszejnov_isothermal_mhd}, but we do not compare it directly here because we do not have a reference version run with exactly the same code version. The most exciting application of this technique to multi-physics simulations may yet be {\it radiation} MHD (RMHD) simulations, whose timesteps are constrained by the light crossing time of a cell, which is almost always orders of magnitude shorter than $\Delta t_\mathrm{grav}$, even if the reduced speed of light approximation is adapted. Our preliminary experiments with large ($2 \times 10^8$ cell) radiation MHD STARFORGE simulations indicate that adaptive force updating can make the cost of gravity totally negligible compared to the cost of the RMHD algorithm, even at very conservative settings ($q_{\rm f} \sim 1/8-1/16$).

\section{Conclusion}
\label{sec:conclusion}
We have proposed a technique for optimising simulations of self-gravitating hydrodynamics in which the gravitational field for gas cells is updated by the gravity solver only as needed, using an inexpensive predictor between updates. We obtained solutions with accuracy comparable to the na\"{i}ve algorithm, but with potentially significant speedups to the gravity calculation, and to the simulations overall if gravity is the dominant cost.

We emphasize that the scheme discussed here is fundamentally different from simply ``sub-cycling" gravity, updating gravity only every fixed number of individual or global hydro time-steps (essentially replacing $\Delta t_{\rm grav}$ with $\Delta t$ in Eq. \ref{eq:updatecondition} and adopting $q_\mathrm{f}>1$). We tie the frequency of updates directly to the updating timescale already deemed necessary for controlling gravitational integration accuracy, which should control the error term adaptively. Simple subcycling is also a viable approach for some problems, but requires more careful tuning by the user because the number of gravity updates required per hydro update is not known {\it a priori} in general.

Rather, the technique explored here can be viewed as a hydro analogue of the method of \citet{ahmad_cohen_scheme}, who accelerated N-body simulations by decomposing the force on a star into rapidly-varying and slowly-varying components, and updated the slowly-varying part less frequently, using a predictor between updates. Here, we take advantage of the separation of timescales between the hydrodynamic evolution at the resolution scale and the gravitational timescale, in analogy to the separation of timescales between the interactions of nearby stars and distant stars in a cluster.

Further gains could potentially be made in some problems by adopting a higher-order force predictor (as opposed to our first-order predictor, Eq. \ref{eq:jerkupdate}), which may require even less-frequent force updates. However the usual trade-off between accuracy and numerical robustness for extrapolation applies: high-order methods will be superior when the underlying variation in $\mathbf{g}$ is guaranteed to be smooth, but may over-fit and give unreliable extrapolations in the presence of noise (e.g. when used with an approximate gravity solver). Hence, such techniques should be tested carefully for a given problem (as always).

There is no such thing as a free lunch: it can easily be reasoned that for a simulation to formally converge to the exact solution, the frequency of force updates {\it must} be increased in conjunction with the spatial and time resolution and the force accuracy. Because $\Delta t_{\rm grav}$ is not necessarily tied to numerical resolution (and is not in our implementation, which takes $\Delta t_\mathrm{grav} = \sqrt{\eta} t_{\rm tidal}$, \citealt{tidaltimestep}), a simulation will not necessarily converge with $q_\mathrm{f}$ held constant because the force error will not approach zero. But this only becomes an issue in the regime where the error introduced by the scheme dominates over other errors in e.g. hydrodynamics, which we have shown not to be the case in at least some instances.

Hence $q_\mathrm{f}$ should be viewed another tunable parameter for the trade-off between accuracy and computational expedience in simulations, alongside e.g. the numerical spatial or mass resolution and the tolerance of approximate gravity solvers. In general, the optimal simulation setup for a given accuracy will lie on a 1D curve within the space of simulation accuracy parameters in which all error terms (including that introduced by our scheme) are comparable. Where exactly this optimal curve lies will generally be problem-dependent: in some problems it may always be advantageous to update gravity every time-step, while in others a large value of $q_\mathrm{f}$ might not noticeably reduce accuracy at all. This additional parameter freedom for optimising simulations may prove useful.

\section{Data availability}
The data supporting the plots within this article and the initial conditions used for numerical tests are available upon request to the corresponding author. A public version of the {\small GIZMO} code is available at \url{http://www.tapir.caltech.edu/~phopkins/Site/GIZMO.html}.

\section*{Acknowledgements}
MYG is supported by a CIERA Postdoctoral Fellowship. This work used computational resources provided by XSEDE allocation AST-190018, the Frontera allocation AST-20019, and additional resources provided by the University of Texas at Austin and the Texas Advanced Computing Center (TACC; http://www.tacc.utexas.edu).




\bibliographystyle{mnras}
\bibliography{bibliography} 




\bsp	
\label{lastpage}
\end{document}